\documentclass[fleqn,twoside]{article}
\usepackage{espcrc2}

\usepackage{graphicx}

\newcommand{\AmS}{{\protect\the\textfont2
  A\kern-.1667em\lower.5ex\hbox{M}\kern-.125emS}}

\title{Large scale simulations of lattice gauge theory: key NSF project of China and recent developments}

\author{Xiang-Qian Luo\address[ZSU]{Department of Physics, Zhongshan University, 
Guangzhou 510275, China}%
        \thanks{Supported by the
Key NSF Project (10235040),  National and Guangdong
Ministries of Education, and Foundation of Zhongshan Univ. Advanced Research Center.}
}
       
\begin{document}

\begin{abstract}
In a recent funded Key Project 
of National Science Foundation of China,
we have planned to do large scale simulations 
of lattice Quantum Chromodynamics, 
using the parallel supercomputing facilities in mainland China. 
Here I briefly review our plan and progress in recent years.
\vspace{1pc}
\end{abstract}

\maketitle

\section{INTRODUCTION}

Lattice gauge theory (LGT) has been accepted as the most powerful tool for investigating non-perturbative strong interactions.
Chinese physicists at Institute of High Energy Physics (IHEP), Institute of Theoretical Physics (ITP) and Peking Univ. (PKU) in Beijing, Nankai Univ. in Tianjin, Chengdu Univ. of Science and Technology and Sichuan Univ. in Chengdu, Zhejiang Univ. (ZJU) in Hangzhou, and Zhongshan Univ. (ZSU) in Guangzhou
have been involved in the study of LGT since early 80's. However, earlier investigations were mainly analytical ones (for a review, see Refs.\cite{guo97,Luo:rv}), due to limited computational facilities.

Thanks to the rapid development of high performance supercomputers in China in late 90's,
success of the Symanzik improvement program, 
and increasing support from NSF (National Science Foundation),
more and more Chinese lattice physicists are doing numerical simulations.

\section{COMPUTING FACILITIES}

There have been many commercial supercomputers installed at research institutions and universities in China. The Beowulf clusters, however, are becoming more and more popular.

The ZSU group built a cluster\cite{Luo:2002kr} of 10 PC type computers. The major difference in our computers from one likely to be found in a home or business
is that each is equipped with two CPUs.  This allows us to roughly double 
our processing power without the extra overhead cost for extra cases, power 
supplies, network cards, etc.  Specifically, we have installed two Pentium 3 processors in each motherboard.  

The Legend Group built a cluster, named DeepComp 1800/LSSC-II, which is installed at Academy of Mathematics and System Science, Beijing, and also open to the Chinese lattice community.
It consists of  Myrinet and 512CPU (Xeon-2GHz). 
Its $Rmax$ and $Rpeak$ are 1.046 Tflops and 2.048 Tflops respectively. 
It was the number 43 of top 500 supercomputers\cite{top500} and the 3rd fastest in Asia in 2002.

\section{NATIONAL SCIENCE FOUNDATION}

The Chinese NSF Committee\cite{nsf} was established in 1986, and it is an organization directly affiliated to the State Council for the management of the National Natural Science Fund. As shown in Tab. \ref{tab1}, these years have seen increasing funding. Each General Project is funded for 100 $\sim$ 200K yuan/3 years (1 Yuan=1/8USD); Each Project for Distinguished Young Scientists is funded  for 0.8M yuan/4 years; Each Key Project is funded for 1 $\sim$ 2M yuan/4 years.

Chinese lattice physicists have got a lot of support from the NSF, including the fund for Distinguished Young Scientists and the Key project\cite{zsu}.

\begin{table*}[htb]
\caption{\label{tab1} Approved funding by the Chinese National Science Fundation (in Million Yuan).}

\begin{tabular}{|c|c|c|c|c|c|c|}\hline

Year & 1996 & 1997 & 1998 & 1999 & 2000	& 2001 \\ \hline

Total &	645.8  & 777.2  & 888.6  & 1083.4 & 1284.3  & 1568.4 \\ 
	
\% of increase/year & 26.4 & 20.4 & 14.3 & 21.9 & 18.5 & 24.5 \\	\hline 

\end{tabular}
\end{table*}

\section{PLAN}

In the recently funded Key NSF Project\cite{zsu}, we have planned to do large scale simulations of lattice QCD, using the parallel supercomputing facilities in China. We plan to develop new numerical methods and study the following hot topics: new hadrons such as glueballs and hybrid mesons, scattering of hadrons, topology of QCD vacuum, transition from the quark confinement phase to quark-gluon plasma phase, quantum instantons and quantum chaos. These investigations will provide useful information for the frontier experiments in particle physics.

\section{RECENT PROGRESS}

\subsection{Improved actions and Hamiltonians}

The idea of Symanzik improvement\cite{Symanzik:1983dc} is to reduce the 
lattice spacing errors by adding new terms to the Wilson actions.
The pursuit of the Symanzik program,
combined with tadpole improvement\cite{Lepage:1992xa},
has recently led to significant progress in LGT, opening the possibility of approaching continuum physics on 
coarse and small lattices. Chinese lattice physicists also contributed actively to the improvement program: the development of the next to nearest neighbor improved quark action\cite{Hamber:1983qa} and improved Hamiltonians for the quarks\cite{Luo:af} and gluons\cite{Luo:1998dx}.

\subsection{Glueballs and hybrid mesons}

Lattice computation of the glueball masses has a long history.
Early investigations used the unimproved 
Wilson gluon action\cite{s6,s26} and Kogut-Susskind gluon  Hamiltonian\cite{Luo:1996ha,Luo:1996sa,Hu:1996ys}. Recent calculations used 
the tadpole improved action\cite{Morningstar:1997ff,s8,s9,Liu:2001wq,MLG} on anisotropic lattices.  In Ref.\cite{s9,Liu:2001wq}, the IHEP group performed a detailed study of the continuum behavior of the glueball $J=0$, $2$, $4$ operators and computed the corresponding spectrum.
In Ref.\cite{MLG}, the ZSU group performed the calculation of the $0^{++}$, $2^{++}$ and 
$1^{+-}$ glueball masses, at $\beta$ = 2.2, 2.4, 2.6, 3.0, and 3.2, 
using the improved gluon action on the anisotropic lattices\cite{Morningstar:1997ff}. 
In comparison with Ref.\cite{Morningstar:1997ff}, we have data
at weaker coupling and larger lattices, 
so that the extrapolation to the continuum limit might be more reliable. 
Our results for the glueball masses
are $M_{G}(0^{++})=1737(\pm52)(\pm107)$, $M_{G}(2^{++})=2409(\pm35)(\pm103)$ 
and $M_{G}(1^{+-})=2946(\pm63)(\pm142)$.

There have been several quenched lattice calculations\cite{Michael} of hybrid meson masses using respectively (1) the Wilson quark action, (2) the SW improved quark action, (3) the NRQCD action, (4) the LBO action, and (5) the improved KS quark action.
In those simulations, configurations were generated using the Wilson gluon action.

In Ref.\cite{Mei:2002ip}, the ZSU group employed {\it both improved gluon and quark actions on
the anisotropic lattice}, which should have smaller systematic errors, 
and should be more efficient in reducing 
the lattice spacing and finite volume effects. 
We obtained $2013 \pm 26 \pm 71$ MeV for the mass of the $1^{-+}$ hybrid meson 
in the light quark sector, 
and  $4369 \pm 37 \pm 99$MeV in the charm quark sector;
the mass splitting between the $1^{-+}$ hybrid meson in the charm quark sector and the 
spin averaged S-wave charmonium was estimated to be $1302 \pm 37 \pm 99$ MeV.

\subsection{QCD at finite chemical potential}

The Hamiltonian formulation of lattice QCD doesn't encounter the complex action problem.
In Refs.\cite{Gregory:1999pm,Fang:2002rk}, the ZSU group developed a Hamiltonian approach at
finite chemical potential $\mu$ and at zero temperature obtained reasonable results in the strong coupling regime.
A first order chiral phase transition was observed at some $\mu_C$. Quantum chaos and quantum instantons\cite{Jirari:1999ij} were also investigated; These phenomena might be relevant for the rich phase structure.

\subsection{Hadron scattering}

The lattice collaboration in Beijing studied scattering of hadrons using tadpole improved clover action on coarse anisotropic lattices\cite{Liu:2001ss,Meng:2003gm}. The
scattering lengths of $\pi \pi$ in the $I$=2 channel and $KN$ in the $I$=1 channel were calculated within quenched approximation. Comparison with the new experiment and Chiral Perturbation Theory was made and good agreements were found.

\subsection{Chiral properties of overlap quarks}

The ZJU group studied the chiral symmetry relation and scaling of the overlap fermions\cite{Dong:2000mr,Ying:aq}.
Divergence of quenched chiral logarithm was found in very light quark regime.

\subsection{Algorithms}

In Ref.\cite{Luo:1996tx}, the ZSU group developed the molecular dynamics equations for doing HMC simulations with dynamical clover quarks and proposed an even-odd preconditioning method for the quark determinant. In Ref.\cite{Jansen:1996yt}, the algorithm was implemented.

The PKU group developed an algorithm for Ginsparg-Wilson quarks\cite{Liu:1998hj}. The ZSU group extended an algorithm for computing the thermodynamical quantities of LGT with Kogut-Susskind quarks, including the chiral limit, to QCD\cite{Luo:2001id}.

The ZSU group developed an algorithm, named MC Hamiltonian\cite{Jirari:1999jn} to study the excited states of LGT and tested in the scalar models\cite{Huang:1999fn}.

Some works mentioned above were done  with the Laval Univ. and Kentucky Univ. groups.

\end{document}